\documentclass[aps,12pt,superscriptaddress,amsfonts,amssymb,amsmath]{revtex4}
\pdfoutput=1

\usepackage{graphicx}
\usepackage{epsfig}
\usepackage{makeidx}
\usepackage{epstopdf}

\begin{document}

\title{Stochastic Effects in a Discretized Kinetic Model\\
of Economic Exchange \footnote{This work was partially supported by the FUB through Research Project DEFENSSE}}

\author{M.L. Bertotti \footnote{Email address: marialetizia.bertotti@unibz.it}}
\affiliation{Free University of Bozen-Bolzano, Faculty of Science and Technology, Bolzano, Italy}
\author{A.K. Chattopadhyay \footnote{Email address: a.k.chattopadhyay@aston.ac.uk}}
\affiliation{Aston University, Nonlinearity and Complexity Research Group \\
Engineering and Applied Science, Birmingham B4 7ET, UK}
\author{G. Modanese \footnote{Email address: giovanni.modanese@unibz.it}}
\affiliation{Free University of Bozen-Bolzano, Faculty of Science and Technology, Bolzano, Italy}

\linespread{0.9}

\begin{abstract}

\bigskip

Linear stochastic models and discretized kinetic theory are two complementary analytical techniques used for the investigation of complex systems of economic interactions. The former employ Langevin equations, with an emphasis on stock trade; the latter is based on systems of ordinary differential equations and is better suited for the description of binary interactions,  taxation and welfare redistribution. We propose a new framework which establishes a connection between the two approaches by 
introducing random fluctuations into
the kinetic model based on Langevin and Fokker-Planck formalisms. Numerical simulations of the resulting model indicate positive correlations between the Gini index and the total wealth, that suggest a growing inequality with increasing income. Further analysis shows, in the presence of a conserved total wealth, a simultaneous decrease in inequality as social mobility increases, in conformity with economic data.
 
\end{abstract}

\maketitle

\section{Introduction}

Microscopic models of economic interactions have been widely studied in the last years. 
The main goal has been an understanding of
how a large number of monetary exchanges among individuals lead to certain income or wealth distributions and to specific values of global indicators like the Gini index or the economic mobility index, see e.g. \cite{Schinkus} which contains a long list of references. In these models the economy of a country is seen as a complex system \cite{Bouchaud,Arthur1,Arthur2} and the Gini and mobility indices are emergent quantities; other macroscopic parameters summarizing the policies of governments, like for instance tax rates, welfare redistribution schemes etc.,
are introduced as input.

Recent extensive work by Piketty and others \cite{Piketty,Milanovic1,Stiglitz,Atkinson,Milanovic2} has highlighted the importance of economic inequality, raising questions about its natural evolution pattern in human societies, and consequently about when and how it is necessary for governments to intervene. Several empirical and theoretical studies have been devoted to the relation between growth and inequality, or to the question, as some put it, whether ``a raising tide lifts all boats''. The answer by Piketty, as is well known, contradicts the view of Kuznets and others \cite{Kuznets,Dollar}.

Silva and Yakovenko \cite{SilvaYak} found in a case study that the fraction of the total population in the Pareto tail of the super-rich changes dramatically when the economy expands or contracts due to external interactions or variations in productivity. Some theoretical models allow to relate the total income of a society to the Gini index of its total income distribution. In \cite{BM1}, it was shown numerically that in a discretized kinetic model including taxation and redistribution, the Gini index of the equilibrium income distribution is an increasing function of the total income defined by the initial condition for a closed system. In a remarkable connection to statistical physics, it was also  found \cite{BM2} that the equilibrium income distribution can be well fitted by the $\kappa$-generalized distribution of Kaniadakis \cite{Kaniadakis1,Kaniadakis2} which displays the same behavior as a function of the temperature and average energy (whereas the Gini index of the Boltzmann-Gibbs distribution is independent of them).

The discretized kinetic model also allows to determine a negative correlation between the Gini index $G$ and economic mobility $M$ which is confirmed by a large body of empirical evidence, nicknamed ``the Great Gatsby law'' \cite{AndrewsLeigh,Corak}. The numerical solutions show \cite{BM33} that when the model parameters (for instance, a parameter $\gamma$ defining a means-tested welfare, and the tax rates $\tau_i$) are varied at total fixed income $\mu$, the corresponding variations of $G$ and $M$ are always of opposite sign, and such that it is possible to trace ``level curves'' for $G$ and $M$ in the $\gamma-\Delta \tau$ plane (with $\Delta \tau = \tau_{n} - \tau_{1}$). Like the relation between $G$ and the total income $\mu$, also this correlation has been established in the model as an equilibrium property.

Experience has also shown, however, that it is impossible to neglect stochastic factors in the evolution of economic systems, especially if strongly influenced by financial markets. As proven, among other instances, by the global crisis of 2008, random fluctuations can generate cascade failures and destabilize a system. The study of these phenomena was pioneered by W.B. Arthur in the 1980s and 1990s \cite{Arthur1,Arthur2}. In conclusion, it is important to include stochastic aspects in the models. Life is full of randomness. While we head to our workplace in the morning, we can estimate the probabilities of many challenges or nuisances which are waiting for us; but we cannot predict those random disruptions that occasionally take us hostage for the whole day...

In this work we model ambient uncertainties by adding stochastic variations of the income distribution to the probability rates of change of the kinetic theory.
In this way we are able to reproduce $\mu/G$ and $M/G$ correlations which match those observed at equilibrium in the absence of noise. We regard this as a confirmation of general statistical properties which appear to apply to individuals in social science as well as to unanimated particles in the physical sciences.

The paper is organized as follows.  In Section \ref{sectionmodel} we briefly recall first the 
evolution equations expressing the two complementary approaches
of linear stochastic models and of discrete kinetic models; this then leads to a novel kinetic Langevin model.
Numerical results for this model are discussed in Section \ref{SectionResults}, where also a Fokker-Planck equation for the case with additive noise is derived.
The last section contains our conclusions and outlook.

\section{Model formulation}
\label{sectionmodel}

\subsection{Linear stochastic vs. kinetic model} 

In the formulation of mathematical models describing an ensemble of interacting agents who exchange money or assets one can choose, among others, 
between the two following schemes:
\begin{enumerate}
\item Consider $N$ agents, with their individual wealth values $w_i$ as fundamental variables; couple each agent to an external random source representing investment or stock trade and possibly also couple each agent to the others with a linear interaction term. The resulting stochastic equations can be solved numerically or transformed, in some cases, into a Fokker-Planck equation. This model, which we call ``linear stochastic model'', has been proposed by Bouchaud and Mezard \cite{BouMez}. Its mean-field approximation has also been independently obtained 
from the stochastic dynamics of a single agent in \cite{Amit}, and applied to an analysis of the poverty index in \cite{AmitPhysica}.
\item Re-group the $N$ agents into $n$ income classes, with $n \ll N$ and define a coupled dynamical system of the Boltzmann type which describes transitions between the classes. The fundamental variables are, in this case, the population fractions $x_i$ ($i$ = 1, 2, ..., $n$) of the classes.  
The interclass interactions are non-linear in these variables
and the evolution equations fit into a discretized kinetic framework \cite{BM1,BM2}.
\end{enumerate}

The two approaches are technically different and lend themselves to the analysis of different issues. 
Random fluctuations and trading are embedded from the start in the linear stochastic model. On the other hand, the discretized kinetic approach allows a more detailed description of the interactions and an analysis of effects like taxation, tax evasion \cite{amc} and welfare redistribution. 
The introduction of a network structure into kinetic theory has been discussed in \cite{BM3} and compared with the analogous structure in the linear stochastic model \cite{Loff}. 
Montecarlo simulations \cite{ab,tos} already constitute a bridge between the linear stochastic and kinetic approaches, because the variables of the simulations are the individual incomes of an ensemble of agents, but their binary interaction rules are of the kinetic type.

\smallskip

We recall next the equations of the linear stochastic model and of the discretized kinetic approach. 

The evolution equations of the linear stochastic model \cite{BouMez,Amit} are
\begin{equation}
\frac{dw_a}{dt}(t) = \eta_a(t) w_a(t) + \sum_{b \neq a} J_{ba} w_b(t) -  \sum_{b \neq a} J_{ab} w_a(t), 
\end{equation}
where $a, b = 1, 2, ... N$, $w_a(t)$ is the wealth of the ${a}^{\text{th}}$ agent, $\eta_a(t)$ is a stochastic noise 
``due to investment in stock markets, housing, etc."
and $J$ is an interaction matrix.
The economic interpretation is that the wealth of each agent varies as a consequence of interactions and stochastic trade, 
the traded amount being proportional to the wealth, and of the same order of magnitude as the income itself.

For the discretized kinetic model, we consider here for simplicity a version 
that does not include tax payment and redistribution. The corresponding evolution equations describe binary exchanges and take the form
\begin{equation}
\frac{{d{x_i}}}{{dt}}(t) = \sum\limits_{h,k = 1}^n {C_{hk}^i{x_h}(t){x_k}(t)} - 
\sum\limits_{h,k = 1}^n {C_{ik}^h{x_i}(t){x_k}(t)},
\label{kin-eq} 
\end{equation}
for $i = 1, 2, ... n$,
where the constant coefficients $C_{hk}^i$, satisfying for any fixed $h$ and $k$
the condition $\displaystyle \sum_{i=1}^n C_{hk}^i = 1$, 
express the probability 
that an individual of the ${h}^{\text{th}}$ class will belong to the 
${i}^{\text{th}}$ class after a direct interaction with an individual of the ${k}^{\text{th}}$ class;
they define 
all the features of the model, as described in detail in \cite{BM1,BM2}, and allow a large degree of flexibility. 
Specifically, the dynamic process
described by the equations (\ref{kin-eq})
is as follows:
a whole of interactions between pairs of individuals occur simultaneously:
for any $h$ and $k$ in $\{1,...,n\}$
individuals belonging 
to the $h$-th income class meet individuals 
of the $k$-th class 
and some money exchange between such pairs takes place.
If at the considered time the fraction of $h$-individuals is $x_{h}$  
and
the fraction of $k$-individuals is $x_{k}$, the number of encounters of these two categories of individuals is the product $x_{h} x_{k}$.
Any single encounter contributes, albeit to a very small extent, to a change in the fraction of individuals in some income classes.
In fact, for each $h \!-\! k$ pair there is 
a probability $p_{h,k} \in [0,1]$ that the $h$-individual will transfer some money to the $k$-th one,
a probability $p_{k,h} \in [0,1]$ that the $k$-individual will transfer some money to the $h$-th one, and also
a probability $1 - p_{h,k} - p_{k,h} \in [0,1] $ that the two do not exchange money.
This is encoded in the coefficients entering in equations (\ref{kin-eq}) which have been proposed in detail in \cite{BM1,BM2}
and can be compactly written as 
\begin{eqnarray}
\label{C}
C_{hk}^i
& = & 
\frac{S}{{\Delta r}}\left[ {{\delta _{h,i + 1}}\left( {1 - {\delta _{kn}}} \right){p_{i + 1,k}} + {\delta _{hi}}\left( {\frac{{\Delta r}}{S} - \left( {1 - {\delta _{in}}} \right)\left( {1 - {\delta _{k1}}} \right){p_{ki}} - \left( {1 - {\delta _{i1}}} \right)\left( {1 - {\delta _{kn}}} \right){p_{ik}}} \right)} \right] \nonumber \\
\ &  &  + \ \frac{S}{{\Delta r}}{\delta _{h,i - 1}}\left( {1 - {\delta _{k1}}} \right){p_{k,i - 1}} ,   
\end{eqnarray} 
where
$S$ denotes a unit of money
(assumed to be much smaller than the difference $\Delta r = r_{i+1} - r_{i}$ between any two consecutive class average incomes) and 
\begin{eqnarray}
\label{p}
{p_{hk}} 
& = & 
\frac{1}{{4n}}\min \{ h,k\}
 \left( {1 - {\delta _{hk}}} \right)\left( {1 - {\delta _{1k}}} \right)\left( {1 - {\delta _{1h}}} \right)\left( {1 - {\delta _{nh}}} \right)\left( {1 - {\delta _{nk}}} \right) \nonumber \\
 \ &  &  + \ \frac{h}{{2n}}{\delta _{hk}}\left( {1 - {\delta _{1k}}} \right)\left( {1 - {\delta _{nk}}} \right) 
 + \frac{1}{{2n}}{\delta _{1k}}\left( {1 - {\delta _{1h}}} \right)\left( {1 - {\delta _{nh}}} \right) \nonumber \\
\ &  &  +\  \frac{k}{{2n}}{\delta _{nh}}\left( {1 - {\delta _{nk}}} \right)\left( {1 - {\delta _{1k}}} \right) 
+ \frac{1}{{2n}}{\delta _{hn}}{\delta _{k1}}.   
\end{eqnarray} 
This is valid for incomes $r_j$ which increase linearly, namely  ${r_j} = j \cdot \Delta r$ (for more general expressions compare Ref.\ \cite{BM1,BM2}). 
The ${\delta _{hk}}$ appearing here denotes the Kronecker's delta and must be defined for indices which go from 0 to $n+1$. 
The same applies for the $p_{hk}$'s: one must extend the definition above to include 
the special cases defined by ${p_{n + 1,k}} = 0$, for any $k$, and ${p_{k,0}} = 0$, for any $k$. 
We also recall here that the meaning of these $p_{h,k}$ is that
typically
poor people pay and receive less than rich people. Special care in the definition of the indices $h, k = 1$ or $n$ accounts for
the impossibility of moving from the first class to a poorer one and from the $n$-th class to a richer one.

The structure of the equations (\ref{kin-eq}) and the choice of the parameters (\ref{C}) and (\ref{p}) guarantees that 
the normalization condition $\displaystyle \sum_{i=1}^n\:x_i(t)=1$
holds true for all $t \ge 0$, provided it holds true for $t=0$.

\subsection{Langevin kinetic model} 

We will now formulate a ``Langevin kinetic model'', i.e., a system of stochastic differential equations whose deterministic part is represented by equations\ (\ref{kin-eq}) while the stochastic part abides Ito statistics (see Risken \cite{Risken}). The choice of Ito statistics, more frequently employed in finance, is suggested by an analogy with the Bouchaud-Mezard model. However, all the numerical results presented in the following are independent from this choice. This leads us to the following form of the model

\begin{equation}
{d x_i} = D_i^{(1)}(x)dt + \sum\limits_{j} D_{ij}^{(2)}(x){\xi _j} \sqrt{\Gamma \:dt},
\label{Langevin}
\end{equation}
where $\Gamma$ denotes the usual noise amplitude and $\xi_i$ denote $n$ independent Gaussian stochastic variables. The deterministic operator is given by

\begin{equation}
D_i^{(1)}(x) = \sum\limits_{h,k} {C_{hk}^i{x_h}{x_k} - \sum\limits_{h,k} {C_{ik}^h{x_i}{x_k}} }.
\end{equation}
For the diffusion term $D_{ij}^{(2)}(x)$, there are several possible choices with a mandatory condition that this term must conserve 
the total population (as recalled above, the deterministic term satisfies this condition: see \cite{BM1,BM2}). This means that 

\begin{equation}
\sum\limits_{i,j}  D_{ij}^{(2)}(x) \xi _j =0
\label{pop-cons}
\end{equation}
has to hold true for any choice of $\{\xi _j\}$. 

The simplest way to obtain this is to start from arbitrary $\{\xi _j\}$ and normalize their sum to zero, 
setting $\xi_i'=\xi_i-(1/n)\sum\limits_{k}\xi_k$, or in the matrix form $\xi_i'=\sum\limits_{j}  D_{ij}^{(2),\text{ADD}}(x) \xi _j$
with
\begin{equation}
D_{ij}^{(2),\text{ADD}}(x) = \left\{ \begin{array}{l}
1 - \frac{1}{n},{\qquad \rm{  if  }}\qquad  i = j\\
 - \frac{1}{n},{\qquad \quad \rm{  if  }}\qquad  i \ne j \, .
 \label{originala}
\end{array} \right. 
\end{equation}
With this choice, the matrix $D^{(2)}$ actually does not depend on $x$ and the noise becomes of the additive kind.

Alternatively, we can introduce a multiplicative noise, such that the random variations in the populations of the classes are proportional to the populations themselves. For this we can first write stochastic variations proportional to $x_i \xi_i$ and then normalize to zero, thus obtaining $\xi_i'=x_i\xi_i-x_i\sum\limits_{k}x_k \xi_k$, or in matrix form
$\xi_i'=\sum\limits_{j}  D_{ij}^{(2),\text{MULT}}(x) \xi _j =0$
with
\begin{equation}
D_{ij}^{(2),\text{MULT}}(x) = \left\{ \begin{array}{l}
x_i\left(1 - x_i \right),{\qquad \rm{  if  }}\qquad  i = j\\
-x_i x_j, {\qquad  \qquad \rm{  if  }}\qquad  i \ne j \, .
\end{array} \right. 
\end{equation}

The interpretation of the stochastic variations is the following: random interactions internal and external to the system produce random transitions in addition to those of the deterministic part of the equations. With additive noise, the random variations for the rich classes are much larger in 
comparison to the class population.
The total income
is not conserved in these fluctuations (compare the numerical solutions given below), so there are also random exchanges with the {\it external world} outside the system.  
We know that the members of an economic system may also interact with the environment outside the system. Examples could be import-export of goods, or incoming-outgoing of tourism, or in investment and stock trading. 

If we want instead to consider a closed system, where the total income $\mu=\sum\limits_{i}r_i x_i$ is conserved, we must impose the further condition
\begin{equation}
\sum\limits_{i,j}  r_i D_{ij}^{(2)}(x) \xi _j = 0 \, .
\label{incomeconserv}
\end{equation}
  
Finding a diffusion matrix which satisfies both conditions (\ref{pop-cons}) and (\ref{incomeconserv})
is not immediate. Nonetheless, it can be achieved by exploiting e.g. the following Proposition 1.
 
\noindent {\bf{Proposition 1}}.
Given $n$ arbitrary variables $z_i$ and $n$ positive constants $r_i$, a linear transformation 
\begin{equation}
{\bar z}_i=z_i+\sum\limits_{j} a_{ij}z_j \, ,
\end{equation}
such that the two conditions
\begin{equation}
\sum\limits_{i} {\bar z}_i=0 \, , \qquad \quad\hbox{and} \quad \qquad \sum\limits_{i} r_i {\bar z}_i=0 
\label{twofirstintegrals}
\end{equation}
hold true, can be defined by taking
\begin{equation}
{a_{ij}} = \frac{{{R_1}({r_i} + {r_j}) - {R_2} - n \, {r_i}{r_j}}}{{n \, R_2 - R_1^2}} \, ,
\label{exprofaij}
\end{equation}
where
\begin{equation}
R_1=\sum\limits_{j} r_j \, , \qquad \quad\hbox{and} \quad \qquad R_2=\sum\limits_{j} r_j^2 \, .
\label{R-equations}
\end{equation}

\noindent
{\it  {Proof:}}
Let a vector $z = (z_1, ... z_n)$ be randomly chosen. Namely, let the $z_i$ with $i = 1, ... n$ be random numbers. 
We define
\begin{equation}
\bar z_i = z_i + \sum_{j=1}^n \, a_{ij} \, z_j \qquad \hbox{for} \ i = 1, ... n.
\label{trlin}
\end{equation}
We want to fix the elements $a_{ij}$ so as to simultaneously enforce the two conditions given in equations (\ref{twofirstintegrals}). Inserting $(\ref{trlin})$ in $(\ref{twofirstintegrals})$ leads us to the following conditions: 
\begin{subequations}
\begin{equation}
\sum_{i=1}^n \big (1+ \sum_{j=1}^n \, a_{ji}\big ) \, z_i = 0 \, , 
\label{|}
\end{equation}
\begin{equation}
\sum_{i=1}^n \big (r_i + \sum_{j=1}^n \, a_{ji} \, r_j\big ) \, z_i = 0 \, .
\label{||}
\end{equation}
\end{subequations}
Since $(\ref{|})$ and $(\ref{||})$ must hold true for any choice of $z_i$, we must have  
\begin{subequations}
\begin{equation}
1+ \sum_{j=1}^n \, a_{ji} = 0 \, , \qquad \hbox{for} \ i = 1, ... n \, , 
\label{|||}
\end{equation}
\begin{equation}
r_i + \sum_{j=1}^n \, a_{ji} \, r_j = 0 \qquad \hbox{for} \ i = 1, ... n \, .
\label{|V}
\end{equation}
\end{subequations}
Equations $(\ref{|||})$ and $(\ref{|V})$ express $2 n$ constraints over $n^2$ elements of $a_{ij}$.
To satisfy these constraints, we seek a minimum of the function involving these $n^2$ variables 
\begin{equation*}
f = \sum_{i=1}^n \sum_{j=1}^n \, a_{ij}^2 = \sum_{i=1}^n \sum_{j=1}^n \, a_{ji}^2 
\end{equation*}
subject to $2 n$-constraints
\begin{eqnarray}
g_i  = & 1+ \sum_{j=1}^n \, a_{ji} & = 0 \, ,   \qquad \hbox{for} \ i = 1, ... n \, , \nonumber \\
h_i  = & r_i + \sum_{j=1}^n \, a_{ji} \, r_j & = 0 \, ,   \qquad \hbox{for} \ i = 1, ... n \nonumber \, .
\label{IIIIII}
\end{eqnarray} 
We introduce the Lagrange multipliers $\lambda_i$ and $\mu_i$ for $i = 1, ... n$, and consider the Lagrangian
\begin{equation*}
L = \sum_{i=1}^n \sum_{j=1}^n \, a_{ji}^2 + \sum_{i=1}^n \lambda_i \, g_i +  \sum_{i=1}^n \mu_i \, h_i \, .
\label{Lagrangian}
\end{equation*} 
The constrained minimum of $f$ has to be found among the critical points of $L$ (as a function of the $a_{ji}$, $\lambda_i$ and $\mu_i$). 
Straightforward calculations lead to the expression given in (\ref{exprofaij}) for $a_{ij}$.
The proposition is then proved.
\hfill {\ensuremath{\blacksquare}}

Therefore we may consider
$D_{ij}^{(2),\text{ADD},\text{Closed}}(x) = \delta_{ij}+a_{ij}.$
In this way, we can introduce an additive noise which, besides the total population, also conserves the total income.

Incorporating a multiplicative noise which conserves $\mu$ requires a more complex procedure
and will be considered in a different paper.

Note that
if only the first, and not the second, of the two conditions in 
$(\ref{twofirstintegrals})$
is required,
then a procedure similar to the one detailed in the proof of Proposition $1$ can be followed, but only one set of Lagrange multipliers $\lambda_i$'s, and not the $\mu_i$'s, will need to be introduced. In that case one gets
$a_{ji} = - \frac{1}{n}, \:\: j = 1, ... n \, , i = 1, ... n$,
which exactly corresponds to the choice relative to the case with non-conserved income and additive noise made in $(\ref{originala})$. 

We conclude this section with a remark concerning class income variables.
Let us define the total income of the class $i$ as $W_i=r_i x_i$. Rewriting the Langevin equation in the variables $W_i$ with $\eta_i$ as their respective noise, a condition analogous to the one given in equation (\ref{pop-cons}) can be obtained:
\begin{equation}
\sum\limits_{i,j}  D_{ij}^{(2)}(W) \eta_j =0 \,.
\label{inc-cons-W}
\end{equation}
This condition implies that the income, not the population, is conserved. If we want to impose the population conservation criterion as well, an extra constraint will be required:
\begin{equation}
\sum\limits_{i,j}  \dfrac{1}{r_i} \:D_{ij}^{(2)}(W) \eta _j =0 \, .
\label{pop-cons-W}
\end{equation}
The two conditions (\ref{inc-cons-W}) and (\ref{pop-cons-W}) can be solved following a similar prescription as laid out generally in Proposition 1.
Therefore the treatment of the kinetic equations with class income variables is equivalent to the treatment with class population variables.

\section{Results}
\label{SectionResults}

\subsection{Numerical solutions}

As a first test of our framework, we computed several numerical solutions of the discretized Langevin equations. 
The construction of the random variables $\xi_i$ is already detailed above. 
Remember that the asymptotic equilibrium state of the deterministic system depends only on the total income $\mu$, 
and not on the initial conditions $x_i(0)$. To improve the convergence of the discretized Langevin algorithm, we let it start from an equilibrium state 
found as the long ran evolution of the classical equations corresponding to an initial condition with some appropriate total income, e.g. for the case of 10 classes, a distribution with $x_3=1$ and all the other $x_i=0$ ($\mu=30$ if $\Delta r$ is taken to be equal to $10$). A reassuring feature of this mathematical structure is the $\Gamma=0$ deterministic limit, where the results all converge to the asymptotic equilibria obtained using the algorithm of \cite{BM1,BM2,BM3} in absence of taxation, as expected.

\begin{figure}[htbp]
\begin{center}
\includegraphics[width=10cm,height=5cm]{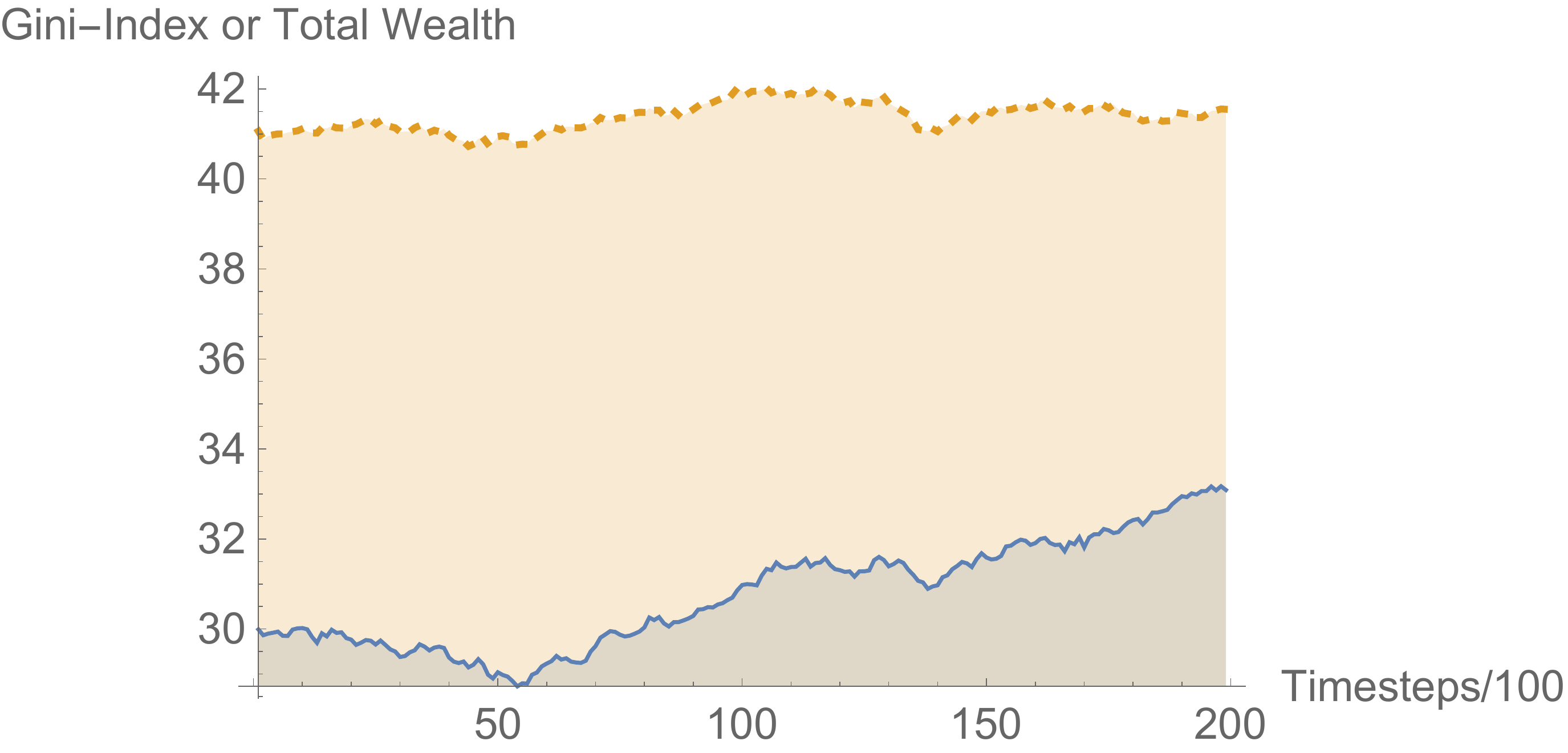}
\caption{Example of a time series (20000 steps) of the total income $\mu$ (solid line) and of the Gini index $G$ (dotted line)  of an income distribution $\{ x_1,...,x_{10} \}$ generated by the Langevin equation (\ref{Langevin}) with additive noise of amplitude $\sqrt{\Gamma}={10}^{-4}$ and $\mu$ not conserved. The value of $G$ has been multiplied by 100. A positive correlation is clearly visible (see discussion in text). The plot shows datapoints after every ${100}^{\text{th}}$ timestep.
} 
\label{fig1}
\end{center}
\end{figure}

The numerical solutions of the Langevin equations measure the fluctuations of the populations $x_i$ of the income classes, in turn leading to the evaluation of  
related quantities like the Gini index $G$, the social mobility $M$ and the total income $\mu$ (when it is not conserved). 
Moreover, it is possible to measure the correlation between $\mu$ and $G$ and also the one between $G$ and $M$.  
The social mobility $M$ is essentially the weighted average, over all classes, of the probability for an individual to be promoted to the upper class in the unit time.
In the present case $M$ is computed using the expression \cite{BM33}
\begin{equation}
M = \frac{S}{{\Delta r}}\frac{1}{{(1 - {x_1} - {x_n})}}\sum\limits_{k = 1}^n {\sum\limits_{i = 2}^{n - 1} {{p_{ki}}{x_k}{x_i}} } \, .
\label{mob}
\end{equation}

In the deterministic system, the dependence of the Gini index on the total income, when the latter is conserved, is monotonic
with a positive slope \cite{BM1}. Also in the present case a clear positive correlation between $\mu$ and $G$ is found (Fig.\ \ref{fig1}). 
In our numerical simulation, we used $\sqrt{\Gamma}\sim{10}^{-4}$ which is the critical order of magnitude of the noise strength 
at which the noise dynamics competes favourably with the system dynamics ($\sqrt{\Gamma}\sim [C]{[x]}^2$).
Note that the deterministic monotone dependence refers to the equilibrium distribution and therefore is verified
through the following steps: a) write initial conditions for the evolution equations with a certain value of the total income $\mu$; 
b) let the system evolve for a long time, towards the asymptotic equilibrium; c) compute $G$ for the resulting income distribution;
d) slightly change the initial conditions and repeat steps a, b, c. On the other hand, in the stochastic equation,
to each random variation of $\mu$ an immediate change in $G$ follows and the correlation is computed based on these changes.

\begin{figure}[htbp]
\begin{center}
\includegraphics[width=10cm,height=5cm]{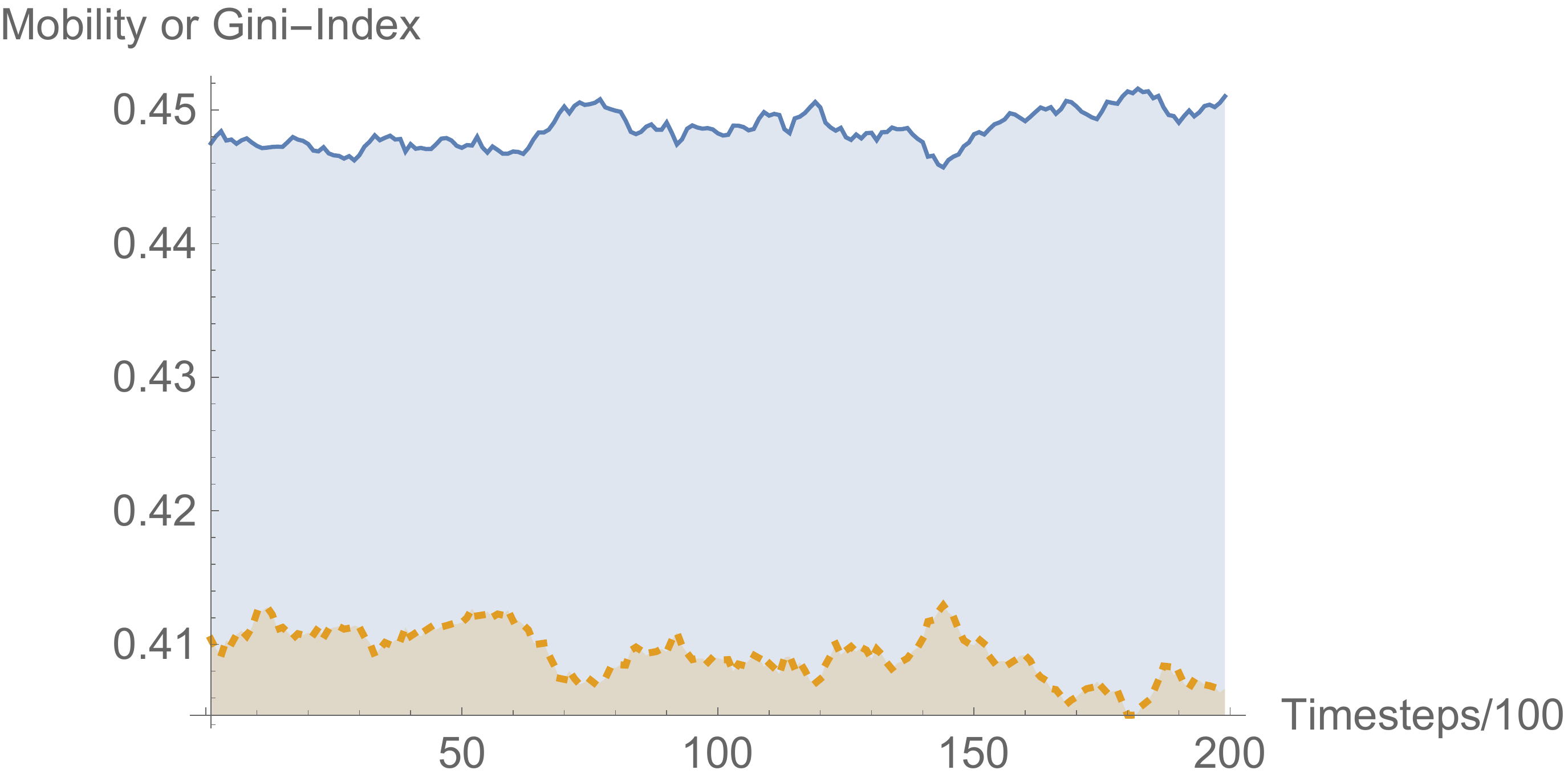}
\caption{Example of a time series (20000 steps) of the social mobility $M$ (solid line) and of the Gini index $G$ (dotted line)  of an income distribution $\{ x_1,...,x_{10} \}$ generated by the Langevin equation (\ref{Langevin}) with additive noise of amplitude $\sqrt{\Gamma}={10}^{-4}$ and $\mu$ conserved. The value of $M$ has been multiplied by 800. A negative correlation is visible (see discussion in text). The plot shows datapoints after every ${100}^{\text{th}}$ timestep.
} 
\label{fig2}
\end{center}  
\end{figure}

Another feature of the deterministic system which was investigated in \cite{BM4} 
in a generalized version including differentiated welfare provision (expressed by a suitable parameter $\gamma$)
and the payment of taxes (related to certain tax rates $\tau_i$) is that, when the model parameters (for instance, the welfare parameter $\gamma$ and the tax rates $\tau_i$) are varied while $\mu$ is kept fixed, 
the corresponding variations of $G$ and $M$ are always of opposite sign and are such that it is possible to trace {\it level curves} for $G$ and $M$ in the 
$\gamma-\Delta \tau$ plane (with $\Delta \tau = \tau_{n} - \tau_{1}$).
In the present model as well a negative correlation between $G$ and $M$ at constant $\mu$ is detected (Fig.\ \ref{fig2}). 
We thus find that certain features of the system's dynamics make their appearance, in a consistent way, both as a consequence on the equilibrium distribution of variations in the parameters 
or as consequences of random internal and external influences represented by the diffusion term of the stochastic differential equations.

\begin{figure}[htbp]
\begin{center}
\includegraphics[width=10cm,height=4cm]{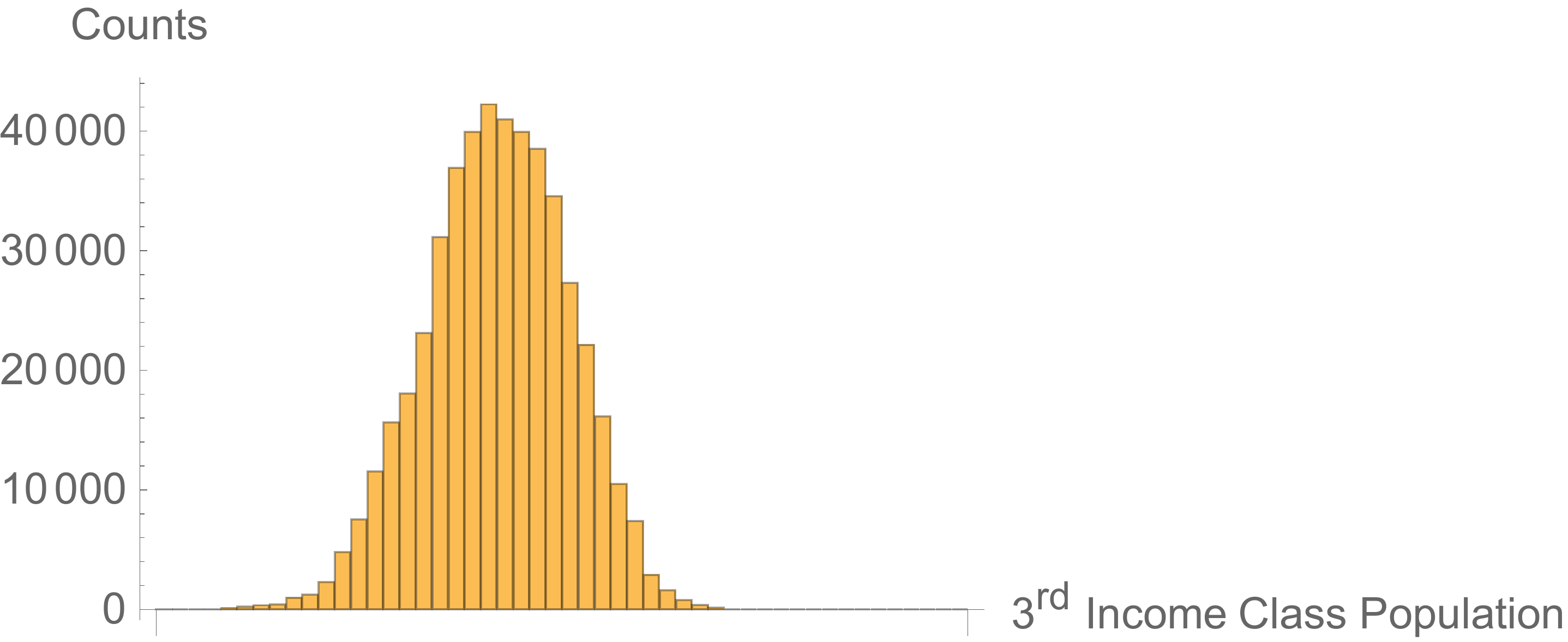}
\caption{Histogram of the statistical distribution of the population $x_3$ of the third income class over 24 realizations, each with 20000 integration steps and $\sqrt{\Gamma}={10}^{-3}$. The noise is such that the total income is conserved. $x_3$ represents the fraction of individuals having income $30$. Each bar has width 0.005, or 0.5\%. 
Values for all variables are given in Table I.
} 
\label{fig3}
\end{center}  
\end{figure}

In order to obtain a complete statistics of the probability distribution of the variables $x_i$ we computed their averages, 
standard deviations and histograms over 24 realizations of the stochastic process, each consisting of 20000 integration steps. 
Fig.\ \ref{fig3} shows the histogram of one of the class populations, namely $x_3$, the fraction of individuals having income 30. 
Naturally, the choice of $x_3$ is only an example, because it is
impossible to represent all these distributions in a single diagram, as could be done for the solutions of the Langevin equations with one income variable \cite{Amit}. This displays one of the interesting aspects of our approach: the populations $\{ x_i, \  i=1,...,n\}$ of the discretized kinetic theory already describe a statistical distribution, to which the insertion of noise adds a further element of realism. 

Table I gives the averages and standard deviations of all the populations, both for the case of non-conserved and conserved total income $\mu$, plus their deterministic values. Note that the standard deviations are larger, in comparison to the population, for the richest classes. This is due to the fact that the noise is additive. It is also apparent from the table that in the case of conserved $\mu$ the deviations from the deterministic values are smaller. This is most probably related to the fact that in the deterministic system all the configurations with the same $\mu$ evolve towards the same equilibrium.

\begin{table}[h]
\begin{center}
\begin{tabular}{|c|c|c|c|c|c|c|c|c|c|c|}
\hline
$i$  & 1 & 2 & 3 & 4 & 5 & 6 & 7 & 8 & 9 & 10  \nonumber \\
\hline
\hline
$x_i$ ($\mu$ non-cons.) &\ 32\ &\ 18\ &\ 11\ &\ 8.1\ &\ 6.4\ &\ 5.6\ &\ 5.2\ &\ 5.0\ &\ 5.1\ &\ 3.7\   \nonumber \\
\hline
$\sigma_i$ ($\mu$ non-cons.) & 8 & 3 & 2 & 2.2 & 2.1 & 2.2 & 2.5 & 2.8 & 3.3 & 3.3   \nonumber \\
\hline
\hline
$x_i$ ($\mu$ conserv.) &\ 37\ &\ 20\ &\ 12\ &\ 8.4\ &\ 6.1\ &\ 4.7\ &\ 3.6\ &\ 3.0\ &\ 3.0\ &\ 2.0\   \nonumber \\
\hline
$\sigma_i$ ($\mu$ conserv.) & 3 & 2 & 2 & 2.2 & 2.1 & 1.8 & 1.6 & 1.4 & 1.4 & 1.3   \nonumber \\
\hline
\hline
$x_i$ (determin.) &\ 37.2\ &\ 19.8\ &\ 12.1\ &\ 8.4\ &\ 6.2\ &\ 4.9\ &\ 3.9\ &\ 3.3\ &\ 2.8\ &\ 1.5\   \nonumber \\
\hline
\hline

\end{tabular}
\end{center}
\caption
{Percentage populations of the income classes and their standard deviations in a case with 10 classes, obtained from the average over 24 realizations, each with 20000 integration steps, of the numerical solutions of the Langevin equations (\ref{Langevin}). In the first two rows: effect of additive noise which does not conserve the total income $\mu$; in this example the average income is larger than the initial value, as can be seen from the diminution of the population in the low-income classes and the increase in the high-income sector, compared to the equilibrium population of the deterministic system (fifth row). In the third and fourth row: effect of additive noise which conserves income.
}
\label{tab1}
\end{table}

\subsection{Fokker-Planck equation with additive noise}

The Langevin model defined in equations (\ref{Langevin}) leads to a Fokker-Planck equation \cite{Risken} which describes the time dynamical evolution of the probability density function of the variables represented in (\ref{Langevin}):
\begin{equation}
\frac{{\partial W}}{{\partial t}} = \left[ { - \sum\limits_i {\frac{\partial }{{\partial {x_i}}}D_i^{(1)}(x) 
+ \sum\limits_{i,j} {\frac{{{\partial ^2}}}{{\partial {x_i}\partial {x_j}}} \left[ D^{(2)}(x) \right]^2_{ij} } } } \right] W.
\label{eqFP} 
\end{equation}

We consider here the matrix $D^{(2)}$ represented in (\ref{originala}), whose elements do not depend on $x$; in this case, 
the second derivatives in (\ref{eqFP}) act directly on $W$. This then leads to
\begin{eqnarray}
\frac{{\partial W}}{{\partial t}} 
&=&  
- \sum\limits_{h,k} {\left[ {\left( {C_{hk}^h{x_k} 
+ C_{hk}^k{x_h}} \right) - \left( {{x_k}\sum\limits_i {C_{ik}^h + C_{kk}^h{x_k}} } \right)} \right]W} \nonumber \\
\ &  &   - \sum\limits_i {D_i^{(1)}(x)\frac{{\partial W}}{{\partial {x_i}}} + \sum\limits_{i,j} {\left[ D^{(2)}(x) \right]^2_{ij}\frac{{{\partial ^2}W}}{{\partial {x_i}\partial {x_j}}}} } .
\label{FPeqn}
\end{eqnarray}

As can be easily noted from the histogram in Fig \ref{fig3}, the steady state probability density function assumes an asymmetric form that makes it imperative that the time varying version of the same too is likely to be asymmetric in nature. All such information is encapsulated in the representation given in equation (\ref{FPeqn}).  In effect, the solution from this model will represent the time variation of the probability density to find the tip of the vector $(x_1,x_2,...,x_n)$ defining the populations of our income classes somewhere in the $n$-dimensional simplex (because the $x_i$ are positive and normalized to 1).
This should be contrasted with some stochastic models used in finance, like the celebrated Black-Scholes model, where the solution of the associated partial differential equation gives essentially a deterministic information, namely the optimal option pricing. 

\section{Conclusion} 

To summarize, in this article, starting from a kinetic model of the Boltzmann kind for economic exchanges, we have explored the role of stochastic uncertainties affecting wealth distribution and the transfer of wealth across income classes. 
In the process, we have formulated income-expenditure dynamics both for conserved and non-conserved total wealth. The analysis relies primarily on direct numerical simulation of the Langevin model alongside qualitative inputs from the Fokker-Planck formulation. Our results indicate that in presence of a Gaussian noise driven dynamics, the equilibrium income distribution is not hugely perturbed. The stochastic fluctuations display clear positive correlations between the Gini-index and the total wealth $\mu$. In presence of a fixed $\mu$, the dynamics shows negative correlation between the Gini-index and economic mobility. In other words, as the total wealth increases, the economic inequality increases whereas in presence of a conserved total wealth, larger social mobility ensures a smaller inequality, or vice versa. For larger or smaller values of noise, the Langevin dynamics is either Brownian in nature or else deterministic. As to the true quantitative implication of the noise distribution in the aforementioned statistics, as also the nature of time dynamics of the probability density function, works are ongoing.

Another issue deserving attention and in fact under investigation is multiplicative noise. The rationale for 
studying it is that one can reasonably expect
the variation of each population fraction to be
proportional to the population fraction itself.
This would prevent relatively large variations 
of the populations of the rich classes, which are typicallly much smaller.

\end{document}